\documentclass{anabs}
\usepackage{graphicx}
\usepackage{times}
\Volume{1-2}              
\Year{2003}              
\Month{02}               
\Pagespan{000}{000}      

\begin{document}
\lhead[\thepage]{H. Ziaeepour \and S. Rosen: Statistical Identification of XMM-Newton Sources Using XID Database}
\rhead[Astron. Nachr./AN~{\bf 324} (2003) 1/2]{\thepage}
\headnote{Astron. Nachr./AN {\bf 324} (2003) 1/2, 000--000}

\title{Statistical Identification of the XMM-Newton Sources Using the XID-DB}

\author{Houri Ziaeepour \and Simon Rosen}
\institute{Mullard Space Science Laboratory (MSSL), Holmbury St.Mary, 
Dorking RH5 6NT, Surrey, UK}


\maketitle

\begin{figure}
\resizebox{\hsize}{!}
{\includegraphics[angle=-90]{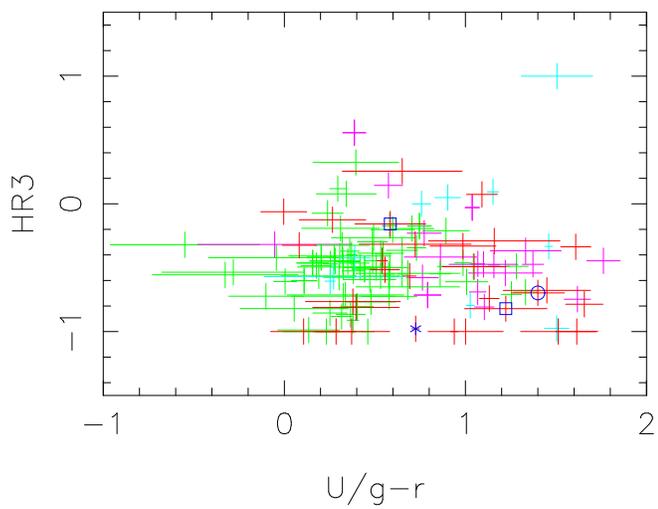}}
\caption{HR3 versus U/g (U or g magnitude), XID-identified objects: 
AGN (green), galaxy (magenta), star (cyan). Unidentified objects from a 
randomly selected XMM field are 
shown in red. A few of these objects have been identified using the Simbad 
catalog (blue markers): circle: AGN of any type, square: galaxy, cross: star.
}
\label{label1}
\end{figure}

The XID Database is the main repository of data for the "XMM-Newton Serendipitous  Survey" (XID-project) and permits easy selection and 
correlation of the XMM X-ray and UV (from Optical Monitor (OM)) data with Optical/IR 
observations. One of the first data-mining applications of the XID database 
is the statistical classification of X-ray sources. Up-to now attempts at  
statistical classification of XMM sources have been mostly 
concentrated on the relation between hardness ratios, fluxes and count 
rates (Della Ceca et al.~2002). Plots of hardness ratio and flux in the 
XID band ($2-4.5 keV$) 
of identified XMM sources however show that there is a large overlap 
between the parameter space of different classes. From ROSAT data it was 
already observed (Carrera et al. ~2001, Page et al.~2000) that 
in color-flux and color-color diagrams, most of identified AGNs and 
unidentified hard sources are segregated from other classes and both 
occupy a relatively small part of the parameter space. The correlation 
between hard X-ray and B-r has been interpreted as being due to 
intrinsic absorption of the X-ray and optical radiation in the source 
(Francis et al. ~1993). We have produced plots of 2-parameter combinations 
of X-ray flux, hardness ratios and optical 
magnitudes $H\alpha-r$, $z-r$, $i-r$, $b-r$, $g-r$ and $U-r$ for 
spectroscopically identified XMM sources. A preliminary 
investigation of $HR2$ and $HR3$ (defined in Osborne ~2000), 
versus $U-r$ and $g-r$ magnitudes 
\footnote {Here we have plotted both 
colours together to increase the 
statistics. Most objects in the database have only one of them. This only 
increases the scatter in $U/g-r$ direction because for most objects in our 
sample the difference between two magnitudes is around 0.7.}
indicates the potential usefulness of these parameters for 
separating different classes of objects. As mentioned above this is the 
result of the intrinsic absorption of X-ray which are re-emitted in longer 
wave lengths. Nonetheless, a more extensive set of 
verified identifications within the database will be needed, specially for 
non-AGN objects, to underpin future analyses. As a first test of this 
method we have plotted detected sources in a randomly 
selected XMM field without any XID identification over the identified 
objects (see Fig.\ref{label1}). Amongst unidentified sources we have found 
5 classified objects from the Simbad Catalog. Regarding the place of these 
objects and other objects of the same class on the plot, it seems that 
classes are insufficiently distinct. Better statistics and greater 
exploration of parameter space are clearly required. 
More specifically, qualitative comparison of a number of plots shows that 
segregation of objects seems more significant in the multiple parameter 
spaces. The study will be extended accordingly. 
Evidently using more sophisticated statistical methods like 
principal component analysis are useful but we postpone the use of 
these methods until we have better statistics. 
Moreover, the database will increasingly 
contain UV data from the OM, yielding a larger 
parameter space for exploitation.

\begin{thebibliography}{}
\bibitem{} Carrera, F.J., Mittaz, J.P.D., Page, M.J., in Proceedings of 
``X-ray astronomy '999:Stellar End-points, AGN and the Diffus Background'', 
eds. G. Malaguti, G. Palumbo \& N. White.
\bibitem{} Della Ceca, R., et al. astro-ph/0202150.
\bibitem{} Francis P.J., et al., 1993, AJ, 106, 417.
\bibitem{} Osborne, J., ed., 2000, SSC-LUX-SP-0004.
\bibitem{} Page, M.J., Mittaz, J.P.D., Carrera, F.J., astro-ph/0103055.

\end{thebibliography}
\end{document}